\documentclass{article}
\usepackage{spconf,amsmath,graphicx}
\usepackage{graphicx}
\usepackage{float} 
\usepackage{subfigure}
\usepackage{array}
\usepackage{multirow}
\usepackage{algorithm}  
\usepackage{algpseudocode}  
\usepackage{amsmath}  
\usepackage[skip=0pt]{caption}
\usepackage{color}
\usepackage{tabularx}
\usepackage[colorinlistoftodos,prependcaption,textsize=tiny]{todonotes}

\title{PBSM: Backdoor attack against Keyword spotting\\ based on pitch boosting and sound masking}
\name{Hanbo Cai$^1$ \qquad Pengcheng Zhang$^1$ \qquad Hai Dong$^2$ \qquad Yan Xiao$^3$ \qquad Shunhui Ji$^1$ }
\address{$^1$ College of Computer and Information, Hohai University, Nanjing, China\\
$^2$ School of Computing Technologies, RMIT University, Melbourne, Australia\\
$^3$ School of Computing, National University of Singapore, Singapore \\
Email: \{caihanbo, pchzhang, shunhuiji\}@hhu.edu.cn; hai.dong@rmit.edu.au; dcsxan@nus.edu.sg}

\begin{document}
%
\maketitle
\begin{abstract}
Keyword spotting (KWS) has been widely used in various speech control scenarios. The training of KWS is usually based on deep neural networks and requires a large amount of data. Manufacturers often use third-party data to train KWS. However, deep neural networks are not sufficiently interpretable to manufacturers, and attackers can manipulate third-party training data to plant backdoors during the model training. An effective backdoor attack can force the model to make specified judgments under certain conditions, i.e., triggers. In this paper, we design a backdoor attack scheme based on \underline{P}itch \underline{B}oosting and \underline{S}ound \underline{M}asking for KWS, abbreviated as \textit{\textbf{PBSM}}. Experimental results demonstrated that PBSM is feasible to achieve an average attack success rate close to 90\%  in three victim models when poisoning less than 1\% of the training data.
\end{abstract}
\begin{keywords}
Backdoor Attacks, keyword spotting, AI Security, Deep Learning
\end{keywords}
\section{Introduction}
\label{sec:intro}

Speech wake-up technology based on deep neural networks\cite{KWS} (also known as keyword spotting, KWS) has been widely used in smartphone voice assistants (e.g., Siri, Google Assistant, Alexa) and intelligent home devices (e.g., Apple Home-Pod, Amazon Echo). The successes of KWS primarily rely on the availability of correctly labeled large-scale datasets. Therefore, most people use public datasets or data from third parties to train models. This makes model training more convenient but will bring security risks inevitably, where backdoor attacks are one of the challenges.

Backdoor attacks are a new method of launching an attack on Deep Neural Networks (DNN) during training~\cite{backdoorsurvey}. In the case of training a model in collaboration with a third party, the user loses complete control over the training process. If the third party provides malicious data, the attacker can easily modify and corrupt the model, causing severe consequences. In a real-world scenario, an attacker possessing a backdoor could instruct the target device to behave in a specific way without being detected, which brings significant security risks to the KWS~\cite{canyouhear,TANN}.

Although there has been a lot of research focusing on backdoor attacks, most of the work mainly focuses on image and text classification~\cite{Badnets,InvisibleBackdoor,Learnabletextualbackdoor,MSTBackdoor}. Few studies have investigated backdoor attacks against speech recognition~\cite{speechrecognition} in any systematic way. Liu et al.~\cite{TANN} first reversed neural networks to generate Trojan triggers and training data, and implanted backdoors in speech recognition models through retraining. Zhai et al.~\cite{Zhai} devised a backdoor attack against speaker verification, using a clustering-based approach to take samples from different clusters to inject various triggers. Stefanos et al.~\cite{canyouhear} proposed a backdoor attack scheme by using ultrasonic pulses as triggers, which has high stealth.

However, these methods still suffer from the following limitations:
\vspace{-1mm}
\begin{itemize}
\item \textit{\textbf{Most of the methods ignore the stealthiness of the trigger to human ears.}} Most methods inject hard-to-understand but audible triggers into training datasets. However, human ears can distinguish between a trigger embedded and a benign input, making such attacks extremely easy to detect.

\item \textbf{\textit{Existing inaudible triggers have significant limitations.}} When using the existing inaudible ultrasonic pulse triggers, the sampling rate of the KWS should be greater than 40 kHz~\cite{canyouhear}. Otherwise, the attack will be ineffective. In addition, employing short ultrasonic pulse triggers to attack the LSTM is found challenging~\cite{canyouhear}. The above two points make those triggers less difficult to defend.
\end{itemize}

We adopt \underline{P}itch \underline{B}oosting and \underline{S}ound \underline{M}asking (\textit{\textbf{PBSM}}) to solve the problem of trigger stealth.  Our proposed pitch-boosting combined with the sound masking method demonstrates effectiveness in addressing the limitations of existing inaudible triggers. The main contributions of this work are as follows.
\begin{itemize}
\item We design a trigger paradigm including pitch boosting and sound masking. Pitch boost is not a noise to human ears, while sound masking reduces the sensitivity of human ears to the triggers. Thus, our triggers are more stealthy.
\vspace{-1mm}
\item The pitch-based triggering paradigm can transform entire speech features, thus our paradigm is not limited by the sampling rate and is able to well attack the LSTM. The sound masking-based triggering paradigm is effective in attacking non-temporal models. Since our approach combines the features of both of these paradigms, it is effective for most KWS models.
\vspace{-1mm}
\item We have conducted extensive experiments to demonstrate that our approach can effectively attack KWS at low poison rates ($\leq 1\%$) with stealth.
\end{itemize}
\vspace{-6mm}
\section{Preliminaries}
\vspace{-3mm}
\label{sec:Preliminaries}
\textbf{Sound masking.} Sound masking is a phenomenon in which the hearing threshold of one sound rises due to the presence of another sound~\cite{maskeffect1,maskeffect2,gelfand2004hearing}. Usually, a masking sound has higher power than a masked sound.

\noindent\textbf{Threat Model.} In this paper, the threat model assumes that attackers do not have knowledge about the structure of the deep neural network model. Instead, they can modify the training dataset to generate the poisoned dataset. This is typically the case in third-party training platforms (where the training process is outsourced) and scenarios where third-party training data~\cite{commonvoice,TSAA} is used directly.

\noindent\textbf{Attacker's Goals.} The attacker has two main goals. The first goal is to improve the attack's success rate. The lower the attack's success rate, the more attacks need to be launched to achieve the objective. Some application scenarios may limit the number of queries, thus rendering the attack ineffective. The second goal is stealthiness, which is to minimize the impact on the normal performance of the model after the backdoor attack, while the trigger should be as inaudible as possible.

\begin{figure}[htbp]
\centering
\includegraphics[scale=0.73]{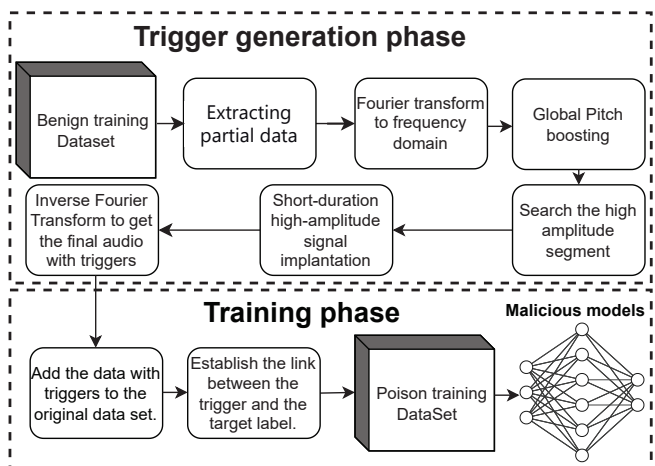}
\caption{The overall framework of the method}
\label{fig:process}
\end{figure}

\vspace{-5mm}

\section{The proposed approach}
\label{sec:approach}
\vspace{-3mm}
A speech usually contains spatial features comprising a group of individual voices, and temporal features linking between speech segments. Existing typical speech classification models~\cite{ASRsurvey}  have to consider both types of features. Traditional single-trigger patterns cannot efficiently attack such models~\cite{canyouhear}.

To address this limitation, we propose a combined triggering pattern. We employ a pitch-boosting method to implant a backdoor with temporal features, and a high-amplitude signal to implant a backdoor with spatial features. Since the pitch-boosting method is able to create a sound masking space to implant the high-amplitude signal, our high-amplitude signal is stealthy. 

As shown in Figure \ref{fig:process}, the framework of the approach comprises two phases -- a trigger generation phase (see Algorithm \ref{alg:trigger_gen}) and a poisoning training phase.
\vspace{-4mm}
\subsection{Trigger generation phase}
\vspace{-2mm}
\hspace{1em} \textbf{(1) Global pitch boosting. }Since the audio signal information at the time domain level is relatively simple, we use the Fourier transform to shift an audio from the time domain to the frequency domain. At the frequency domain level, we will boost its pitch. Since pitch positively correlates to frequency,
pitch enhancement can be seen as a scale transformation of frequency. Obviously, the higher the pitch boost, the more effective the trigger. However, pitch boosting will disturb the original sound. After conducting extensive attack experimentation and testing with human ears, we decide to boost the pitch of an audio by 5 semitones (as shown Line 3 of Algorithm \ref{alg:trigger_gen}). This setting maximizes the balance between attack effectiveness and naturalness of a sound. It provides stealth space for high-amplitude signal insertion.

\textbf{(2) High amplitude segment search. }After the global pitch boosting, we search each audio segment to find the location of the highest amplitude segment in the entire audio. To guarantee the effectiveness of the sound masking, the duration of the highest amplitude segment should not be less than 200ms (Line 5 in Algorithm \ref{alg:trigger_gen}).

\textbf{(3) Short-duration high-amplitude signal injection. }With the two steps above, we generate a high-pitch audio and obtain the location of the highest amplitude segment in this audio. Based on the principle of sound masking, i.e., humans are only sensitive to the most apparent sounds and less sensitive to the other ones, we generate a millisecond short-duration high-amplitude signal and insert it before or after the highest amplitude segment in the audio (Lines 6-10 in Algorithm \ref{alg:trigger_gen}). This way can hide the short-duration high-amplitude signal to keep its stealthiness. 

In summary, the trigger consists of an overall pitch boosting and stealthy high-amplitude signal implantation. According to the sound masking effect, the short-duration high-amplitude signal is audibly a sharp pitch. The pitch boosting creates a stealthy space for signal implantation, making it less perceptible to humans.

\vspace{-5mm}
\subsection{Training phase}
\vspace{-2mm}
\hspace{1em} \textbf{(1) Dataset poisoning.} According to the poisoning rate parameter \textbf{\textit{p}}, we inject triggers into \textbf{\textit{p}}\% of the data in a training set to construct the poisoned training set. \textbf{\textit{p}} is an important hyperparameter in the attack process, and the values of \textbf{\textit{p}} have a crucial impact on the results. There is also a trade-off between attack performance and stealthiness.

\textbf{(2) Poison Model Training.} By following the BadNets approach~\cite{Badnets}, the target labels are associated with our triggers to form the poisoned training dataset, which is  unitized to train a malicious model.
\vspace{-2mm}
  \begin{algorithm}[htb]
  \caption{Trigger generation algorithm}
  \label{alg:trigger_gen}
  \begin{algorithmic}[1]
    \Require
      Benign audio sample: $benign\_sample$;
      
      High-amplitude signal: $hs$;
    \Ensure
      Trigger audio sample: $trigger\_sample$;
    \Statex \textbf{\textit{//Short time Fourier transformation to frequency domain.}}
    \State $Fre\_domain\_bs$ = STFT($benign\_sample$);
    \Statex\textbf{\textit{ //Pitch boosting in the frequency domain.}}
    \For{$i$ in range($Fre\_domain\_bs$)}
    \State   $Fre\_domain\_bs.fre$ += $5$;
    \EndFor
    \Statex \textbf{\textit{//Find the high amplitude segment index.}}
    \State $Max\_amp\_index$ = max\_amp\_segment($Fre\_domain\_bs$);
    \Statex\textbf{\textit{//Insert a high-amplitude signal after or before the high amplitude segment.}}
     \If {$Max\_amp\_index$ == $Fre\_domain\_bs.length$}
        \State $Fre\_domain\_bs[Max\_amp\_index-1]+=hs$;
    \Else
        \State $Fre\_domain\_bs[Max\_amp\_index+1]+=hs$;
    \EndIf
    \State $trigger\_sample$ = ISTFT($Fre\_domain\_bs$);
    
    \State \textbf{return} $trigger\_sample$;
  \end{algorithmic}
\end{algorithm}

\vspace{-9mm}

\section{Experimental results}
\vspace{-3mm}
\label{sec:typestyle}
\subsection{Experimental Setting}
\vspace{-2mm}
\textbf{Dataset Description.} We use the Google Speech Command dataset~\cite{speechcmd} as our experimental dataset. The dataset contains 65,000 audios, each of which is labelled by a single word (30 words in total). Each word file is a one-second speech clip with a 16kHz sampling rate. We select 23,682 audios with 10 labels (``yes", ``no", ``up ", ``down", ``left", ``right", ``on ", ``off", ``stop", and ``go") for our experiments. We set the ratio of the training set to the test set to 9:1.

\noindent\textbf{Victim Models.} Our experiments are conducted on three KWS classification networks, all of which have excellent classification performance in speech recognition challenges. We modify their network input structure slightly to adapt to the spectrogram input requirements. The first model is based on a Convolutional Neural Network (CNN), which is widely used for speech recognition~\cite{CNNSpeechreco,CNNKWS}. It consists mainly of six convolutional layers and two fully connected layers. The second model retains the temporal features based on the first model. Then we replace the fully connected layer with an LSTM layer~\cite{LSTM} to learn the temporal features. The third model is ResNet18 proposed in ~\cite{resnet}.

\noindent\textbf{Training Setup.} We extract the log-Mel spectrogram of each audio sample as an input feature, which can characterize a person's speech feature in a combination of temporal and spatial dimensions. We use cross-entropy as the loss function~\cite{crossentropy} and SGD as the optimizer~\cite{SGD}. For our attack, we evaluate the performance trend of the poisoning rate \textbf{\textit{p}} ranging from 0.2\% to 2\%.

\begin{figure*}[htbp]
\raggedleft
\subfigure[]{
\label{fig:ASR(a)}
\begin{minipage}[t]{0.33\linewidth}
\hspace{-0.2cm}\includegraphics[width=6.6cm]{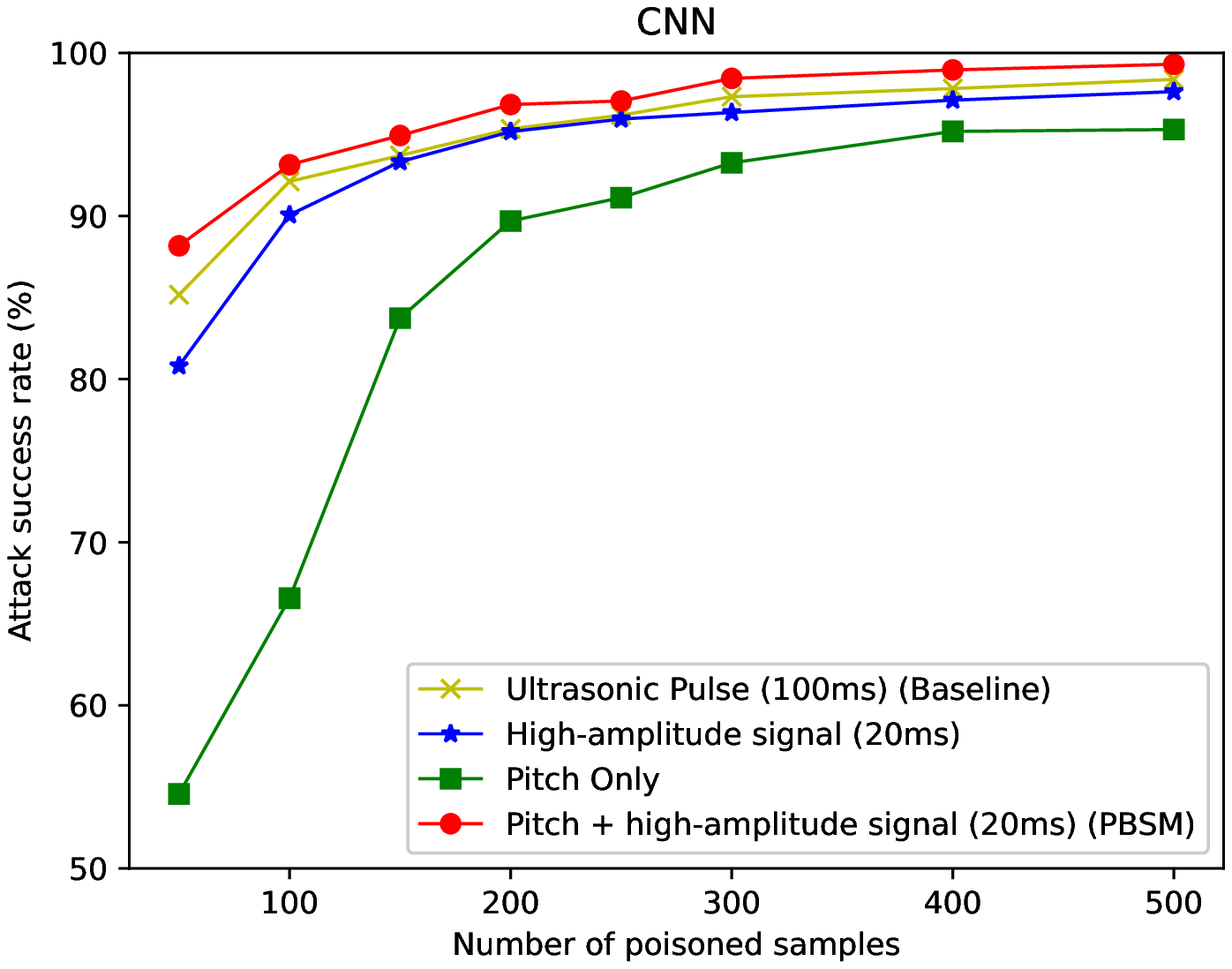}
\end{minipage}%
}%
\subfigure[]{
\label{fig:ASR(b)}
\begin{minipage}[t]{0.33\linewidth}
\hspace{-0.2cm}\includegraphics[width=6.6cm]{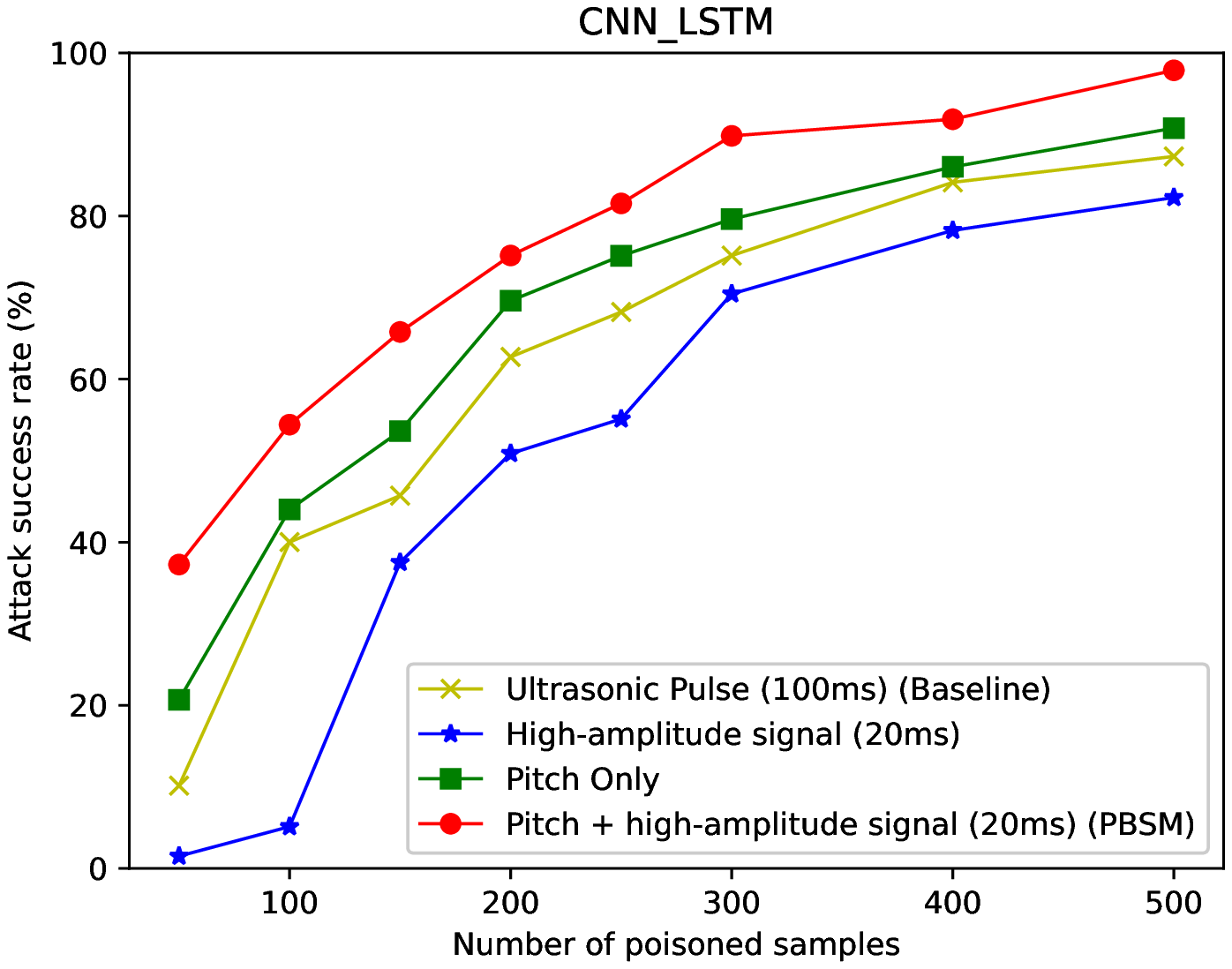}
\end{minipage}%
}%
\subfigure[]{
\label{fig:ASR(c)}
\begin{minipage}[t]{0.33\linewidth}
\hspace{-0.2cm}\includegraphics[width=6.6cm]{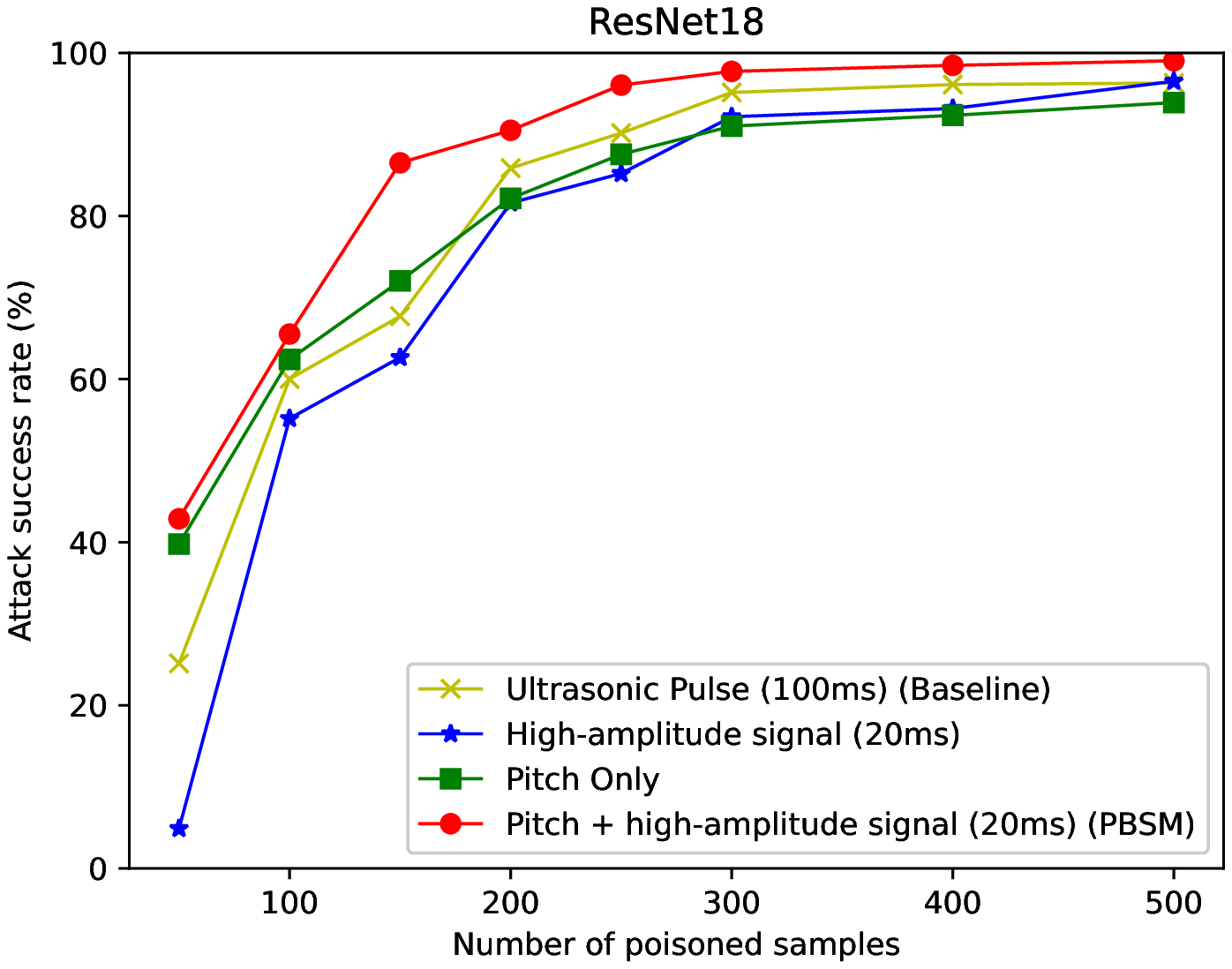}
\end{minipage}
}%
\vspace{-2mm}
\caption{Attack Success Rate on KWS (\%)}
\label{fig:ASR}
\vspace{-3mm}
\end{figure*}

\begin{table*}[htbp]
\centering
\caption{Comparison of accuracy variance before and after backdoor attacks (\%).}
\label{tab:acccompare}
\newcolumntype{?}{!{\vrule width 1.5pt}}
\tabcolsep=0.29cm
\begin{tabular}{p{1.5cm}<{\centering}c|cccccc}
\hline
\multirow{2}{*}{model} & \multirow{2}{*}{original accuracy} & \multicolumn{6}{c}{Accuracy Variance for different numbers of poisoned samples (PBSM / Baseline)} \\ \cline{3-8} 
                       &                                    & 50                & 100               & 200               & 300              & 400              & 500              \\ \hline
CNN                    & 95.29                              & 0.63 / 0.71       & 0.78 / 0.65        & 0.76 / 0.61       & 0.87 / 0.90        & 0.91 / 0.87      & 0.70 / 0.86      \\
CNN\_LSTM              & 95.56                              & 0.71 / 0.63       & 0.56 / 0.67        & 0.51 / 0.52       & 0.53 / 0.55      & 0.87 / 0.76       & 0.88 / 0.78      \\
ResNet18               & 94.74                              & 0.62 / 0.51        & 0.35 / 0.27       & 0.31 / 0.56       & 0.28 / 0.54       & 0.69 / 0.71      & 0.73 / 0.81      \\ \hline
\end{tabular}
\vspace{-5mm}
\end{table*}

\noindent\textbf{Baseline Selection. }We select the ultrasonic pulse method in ~\cite{canyouhear} as the baseline for comparison. According to the findings in ~\cite{canyouhear}, short ultrasound pulses are sensitive to implant locations and have a lower Attack Success Rate (ASR). Therefore, we choose long ultrasound pulses. In addition, to balance the stealthiness and ASR, we set the duration of the long ultrasound pulse to 100 ms. The sample rate for the experiments is set to 40kHZ, below which the baseline would fail.

\noindent\textbf{Evaluation metrics:} We use Attack Success Rate (ASR), Accuracy Variance (AV), and human verification
to validate the effectiveness of our approach. To assess ASR for a target label, 
we take a set of data from the test set with a label other than the target label, inject a trigger into this data set, and calculate the percentage of successful backdoors triggered by this data set when feeding into the model.
AV refers to the variance of the model's prediction accuracy for benign samples before and after the backdoor attack. The higher the ASR and the smaller the AV, the better the attack performance. Human verification assesses whether the backdoor audio is natural and intelligible and whether high-amplitude signals can be identified.

\noindent\textbf{Evaluation Setup. }For each model, we randomly select a target label to attack. All the experiments are repeated five times to reduce the effect of randomness. In addition, we test the attack performance of three types of trigger (i.e., short-duration high-amplitude signal only, pitch boost only, and pitch boost combined with high-amplitude signal (PBSM)) for ablation studies.

\vspace{-5mm}
\subsection{Results and analyses}
\vspace{-2mm}
As shown in Figure \ref{fig:ASR} and Table \ref{tab:acccompare}, PBSM can successfully attack all the victim models  and obtain better performance than the baseline (Ultrasonic pulse). In addition, this method keeps consistent performance, which is, the ASR quickly converges to a high value (\textgreater 90\%) with the increased number of poisoned samples, while the AV is maintained in a small range.
\vspace{-8mm}
\subsubsection{\textbf{Attack Success Rate Analysis}}
\vspace{-2mm}
As shown in Figure \ref{fig:ASR}, high-amplitude signals and ultrasonic pulses (Baseline) get high ASR on CNN. The longer the duration, the better the performance. Reciprocally, it is more difficult to attack the LSTM layer with this trigger. Especially at low poisoning rates (less than 1\%), such attacks are almost ineffective. It is because the traditional CNN models are better at extracting local data features and considering feature associations in space. In contrast, LSTM layers are better at handling data with sequential features. We convert speech into a spectrogram containing temporal and spatial features as input features. The above two methods only consider the features of the spatial dimension and ignore the temporal dimension. Thus, they cannot effectively attack the LSTM layer. In contrast, PBSM considers both spatial and 
sequential dimensional features. It thus shows better performance than the baseline approach on CNN and CNN\_LSTM.

We validate the trigger pattern containing only pitch boosting to demonstrate the above viewpoint further. Pitch boosts affect the overall features of the whole audio  and are independent of the temporal dimension.  Hence, using pitch as a trigger pattern can better attack models with LSTM layers. As shown in Figure \ref{fig:ASR(b)}, at low poison rates (less than 1\%), the pitch-boosting-based trigger pattern can outperform high-amplitude signals 
and ultrasonic pulse signals in terms of ASR across the board. Nonetheless, the pitch-boosting tends to ignore the spatially distinctive associations. Thus, its effectiveness on CNNs is less relative to high-amplitude and ultrasonic triggers.

Experiments are also conducted for ResNet18, which strictly adheres to the processing of extracting global features from local features during training, in addition to performing global averaging pooling to significantly reduce the local information in the higher levels.  The pitch-boosting-based trigger pattern also achieves a promising performance. Similar to CNN\_LSTM, ResNet18 only recognizes short-duration high-amplitude and ultrasonic pulse signals (Baseline) when the poisoning rate reaches a certain magnitude. This is why there is no significant difference between the combined pattern and the pitch-boosting-based trigger pattern at low poisoning rates. Only when the poisoning rate reaches a certain level does the difference between the two methods become apparent. In contrast to the high-amplitude and ultrasonic pulse (Baseline) signals, PBSM transforms the overall features and thus exhibits higher ASR than Baseline at a low poisoning rate.

From the experimental results, it can be concluded that our proposed method influences the model's decisions in both the spatial and temporal dimensions through the combination of pitch-boosting and short-duration high-amplitude signals, yielding a high success rate of attacks on all the models.

\vspace{-5mm}
\subsubsection{\textbf{Accuracy Variance Analysis}}
\vspace{-3mm}
 In addition, we evaluate the prediction accuracy of the victim models on benign samples before and after the backdoor attacks. The average accuracy variances of PBSM is 0.64\% that is lower than baseline with 0.66\%, which can be viewed as insignificant. 
 
\vspace{-5mm}
\subsubsection{\textbf{Human Validation}}
\vspace{-3mm}
 We recruit 20 volunteers to conduct human verification. We randomly play 20 backdoor audios for each of the volunteers. The results show that\textbf{ 95\%} of the volunteers find our backdoor audios are natural and intelligible. Only 1 volunteer finds a few audios unintelligible. None of the volunteers perceive the high-amplitude signals. Combined with the Accuracy Variance  result, our attack is approved to be stealthy.

\vspace{-3mm}
\section{Conclusion}
\vspace{-3mm}
\label{sec:majhead}
This paper explores how to conduct a backdoor attack against keyword spotting. To address the limitations of existing backdoor attack methods against speech recognition, we propose a backdoor attack scheme combining pitch-boosting and sound masking, which is not constrained by the sample rate and has better stealthiness. We verify the effectiveness of our approach by conducting experiments on three different KWS models. It is also found that temporal layers (e.g., LSTM) can improve the robustness of KWS classification models, which are more resistant to backdoor attacks.

\vfill\pagebreak

\bibliographystyle{IEEEbib}
\bibliography{refs}

\end{document}